# Low temperature annealing studies of $Ga_{1-x}Mn_xAs$

I. KURYLISZYN [1], T. WOJTOWICZ [2], X. LIU [2], J.K FURDYNA [2], W. DOBROWOLSKI [1],

[1] Institute of Physics, Polish Academy of Sciences, Warsaw, Poland

[2] Department of Physics, University of Notre Dame, Notre Dame, USA

J.-M. BROTO [3], M. GOIRAN [3], O. PORTUGALL [3], H. RAKOTO [3], B. RAQUET [3]

[3] Laboratoire National des Champs Magnetiques Pulses Toulouse, France

High- and low-field magneto-transport measurements, as well as SQUID measurements of magnetization, were carried out on $Ga_{1-x}Mn_xAs$ epilayers grown by low temperature molecular beam epitaxy, and subsequently annealed under various conditions. We observe a large enhancement of ferromagnetism when the samples are annealed at an optimal temperature, typically about $280^0$C. Such optimal annealing leads to an increase of Curie temperature, accompanied by an increase of both the conductivity and the saturation magnetization. A decrease of the coercive field and of magnetoresistivity is also observed for $Ga_{1-x}Mn_xAs$ annealed at optimal conditions. We suggest that the experimental results reported in this paper are related to changes in the domain structure of $Ga_{1-x}Mn_xAs$.

## 1. INTRODUCTION

Ferromagnetic semiconductors based on III-V compounds continue to be the subject of intense interest due to their important new physical properties, as well as to the possibility of spin-electronic device applications [1,2]. In particular, the physics of carrier-induced ferromagnetism observed in $Ga_{1-x}Mn_xAs$ has become the focus of great current interest. It is widely accepted in this context that ferromagnetic coupling of Mn ions is mediated by free holes (see [3] and references therein).

Ferromagnetic ordering was reported for $Ga_{1-x}Mn_xAs$ grown by low temperature molecular beam epitaxy (LT-MBE) for $x > 0.02$, with Curie temperatures as high as 110K [4]. These observations are consistent with the Zener model of ferromagnetism proposed by Dietl *et al.* [5]. Recently, Jungwirth *et al.* [6] presented theoretical calculations of the Curie temperature in $Ga_{1-x}Mn_xAs$ based on a model where the local Mn moments ($S$=5/2) are exchange-coupled to itinerant holes in the valence band of semiconductor host. This model (which goes beyond the standard mean-field theory) predicts an enhancement of $T_C$ in $Ga_{1-x}Mn_xAs$ due to the exchange and correlation with the itinerant hole system, as well as a suppression of $T_C$ due to collective fluctuations of the ordered moments. The Curie temperatures $T_C^{est}$ estimated by this model are in good agreement with the experimentally observed values of the transition temperature, and remain very close to the mean-field values $T_C^{MF}$, justifying the mean-field description of ferromagnetism in this material.

It has recently been established that the Curie temperature of as-grown GaMnAs epilayers can be further improved by heat treatment (low temperature annealing) [7,8,9,10]. In the present study we investigate the effects of such low temperature annealing on the electronic and magnetic properties of $Ga_{1-x}Mn_xAs$. The purpose of this paper is to present the results of magnetic and magnetotransport measurements performed in the range of low and high magnetic fields (up to 30T) on annealed and as-grown samples in a wide range of Mn concentrations ($0.01 < x < 0.084$), in order to show how post-growth heat treatment affects the physical properties of $Ga_{1-x}Mn_xAs$.

**2. EXPERIMENT**

The $Ga_{1-x}Mn_xAs$ epilayers investigated in this paper were grown by LT-MBE on (100) GaAs semi-insulating substrates at temperatures in the range between $T_S \sim 250^0$C – $275^0$C. The details of the growth method were described elsewhere [9]. The thicknesses of the epilayers ranged between 105 nm and 302 nm. The Mn concentration was determined using two different methods. First, the Mn content was estimated from the change in the growth rate monitored by RHEED oscillations. Second, the lattice constants of $Ga_{1-x}Mn_xAs$ epilayers were measured using x-ray diffraction (XRD), and the Mn concentration was determined by assuming that the GaMnAs layer is fully strained by the GaAs substrate, and that it obeys Vegard's law.

The as-grown samples were then cleaved into a series of pieces for systematic annealing. The samples were annealed at low temperatures in the range between $260^0$C and $280^0$C, under a fixed flow of $N_2$ gas of 1.5 SCFH (standard cubic feet per hour). The annealed samples were characterized by means of zero-field resistivity and SQUID measurements, allowing us to pinpoint the optimal annealing conditions. The results of low magnetic field transport and magnetic measurements are described in detail in Ref. [9]. Briefly, we observe that annealing leads to large changes in both transport and magnetic properties for a narrow window of annealing temperatures (close to the LT-MBE growth temperature). For $Ga_{1-x}Mn_xAs$ specimens in the wide range of Mn concentrations investigated in this paper ($0.01 < x < 0.084$) the optimal annealing temperature $T_a$ was found to be around $280^0$C. Such optimal annealing conditions ($T_a = 280^0$C, $t$ = 1h) are observed to lead to an increase of both the Curie temperature and the conductivity. For example, for $x = 0.083$ optimal annealing shifted the value of $T_\rho$ from 88K for the as-grown sample to 127K, whereas annealing at a higher temperature ($T_a$=$350^0$C) has resulted in a lower Curie temperature ($T_\rho$=30K). Here $T_\rho$ denotes the temperature at which the zero-field resistivity shows a cusp, a point that closely coincides with the Curie temperature as measured by magnetization. In general these large changes in the Curie temperature of $Ga_{1-x}Mn_xAs$ observed after optimal annealing are accompanied by large changes in the conductivity of the material. We found, however, that for $Ga_{1-x}Mn_xAs$ with low Mn concentrations, $x < 0.05$, the influence of annealing on both the Curie temperature and on the conductivity is weak.

Determination of the free carrier concentration of ferromagnetic $Ga_{1-x}Mn_xAs$ epilayers from Hall measurements is a difficult task because, in addition to the normal term proportional to the magnetic field, the Hall resistivity now contains a supplementary term proportional to the magnetization (called the anomalous Hall resistivity, AHE). The presence of AHE (which dominates the Hall resistivity at typical laboratory fields and temperatures) thus complicates the evaluation of the free hole concentration from Hall measurments. In principle this difficulty can be circumvented when the Hall measurements are performed at low temperatures and at magnetic fields sufficient high to saturate the magnetization.

We measured the Hall resistivity and conductivity simultaneously in high pulsed magnetic fields (up to 30T) for the series of as-grown and annealed epilayers of GaMnAs. Figure 1 shows typical anomalous Hall voltage signals as a function of magnetic field for two samples of GaMnAs with $x = 0.083$: as-grown, and annealed at an optimal temperature ($289^0$C). Note that the majority of the Hall voltage comes from the anomalous Hall effect. In the range of very high magnetic fields, where one may expect the magnetization to be saturated, the slope of the Hall voltage should be related to the free carrier concentration. In our experiments the difference between the slopes of the Hall voltage in high magnetic fields is very small, and the present data (measured up to 30T) do not give a unique value for the hole concentration of the investigated epilayers. Contrary to the expectation that the free hole concentration should increase after optimal annealing, as shown by earlier results obtained by electrochemical capacitance voltage (ECV) profiling [10], the data reported here appear to indicate a decrease of the free carrier concentration. In our opinion, however, the slope seen in our experimental data reflects a dependence of $M$ on $B$ persists to these high fields, rather then the "normal" Hall effect behavior. This suggests that 30T is not sufficient to saturate the AHE contribution, so that the Hall and resistivity measurements should be performed at still higher in order fields to determine the free carrier concentration of GaMnAs.

We measured the conductivity in high magnetic fields with the field applied perpendicular to the plane of the film, and found that annealing leads to a very pronounced decrease of the magnetoresistivity. To illustrate this, in Fig. 2 we show the magnetoresistivity as a function of magnetic field measured at low temperatures up to 30 T for epilayers with high Mn concentration ($x = 0.081$, $x = 0.084$). All samples show a negative magnetoresistance (*MR*) in the range of high magnetic fields. It is clear from fig. 2 that annealing at optimal conditions leads to a significant decrease of *MR*. In low magnetic fields the *MR* curves exhibit hysteretic behavior. This effect is shown at Fig. 3.

We noted also that the large increase of magnetoresistivity is accompanied by changes of magnetization measured as a function of magnetic field *M(B)* and temperature *M(T)*. The hysteresis loops were investigated using a SQUID magnetometer at low temperatures, with the magnetic field applied in the plane of the film. Figure 4 shows that annealing at the optimal temperature leads to a decrease of the coercive field of the hysteresis loop. A clear increase of the saturation magnetization $M_S$ is also observed after heat treatment at optimal conditions, indicating that annealing increases the concentration of magnetically-active Mn ions. As temperature is reduced below the Curie temperature, we observe an increase in the coercive field (not shown) that correlates with the hysteretic behavior in the *MR* measurments. We note parenthetically that the low-field hysteretic feature of the magnetoresistivity becomes increasingly pronounced as the temperature decreases.

## 3. DISCUSSION AND CONCLUSIONS

We have investigated the electrical and magnetic properties of as-grown and annealed GaMnAs epilayers. We have shown that annealing at optimal conditions leads to a strong enhancement of ferromagnetism in GaMnAs, in the form of an increase of the Curie temperature accompanied by an increase of conductivity and saturation magnetization $M_s$. We attribute such large changes in the transport and magnetic properties to annealing-induced lattice rearrangement of the positions of the Mn atoms in the GaMnAs lattice [10], as discussed below. One of the high-Mn concentration samples from this investigation ($x$ ~ 0.08) was also characterized by channeling Rutherford backscattering (c-RBS) and channeling particle induced X-ray emission (c-PIXE). The results are presented elsewhere [10]. Briefly, the channeling experiments revealed that the large annealing-induced increase of $T_C$, accompanied by an increase of saturation magnetization and of free carrier concentration, can be attributed to the relocation of Mn atoms from interstitial ($Mn_I$) to substitutional sites ($Mn_{Ga}$) or to random positions (which Mn can occupy if it precipitates in the form of other phases, e.g., as MnAs). The removal of a significant fraction of highly unstable interstitial Mn ($Mn_I$) provides a clear explanation of the three effects observed after annealing, i.e., the increase of the hole concentration, the increase of the Curie temperature, and the increase in the saturation magnetization observed at low temperatures [10].

Finally, the LT annealing is also seen to affect the magnetoresistivity and hysteresis loops. Specifically, a pronounced decrease of the coercive field is observed in the annealed samples. This is accompanied by a low-field hysteretic behavior of the *MR* data, as seen in Fig. 3. A clear reduction of the negative magnetoresistivity is also observed at high fields in samples that were annealed under optimal conditions (see Fig. 2).

At present one can only present a speculative and phenomenological interpretation of the results just described. Recently, Fukumura *et al.* [11] showed that the films of GaMnAs with in-plane magnetization have an unconventional domain structure, showing a random arrangement of domains. It is very likely that the observed hysteresis loop features – the value of the coercive field, the shape of the loop, and saturation magnetization - are determined by the domain structure of the material. It is well known that transport properties of ferromagnetic materials can be modified by the details of the domain walls; in particular, the domain wall structure can increase or decrease the resistance of the system [12]. Moreover, it was recently shown that the shape and the positions of the *MR* peaks depend on the domain structure and on the "squareness" of the hysteresis loops [13]. We therefore suggest that the observed pronounced changes in both the *MR* and in magnetization are related to the changes of the domain structure in the GaMnAs system.

**ACKNOWLEDGMENTS**

This work was partially supported by the European Community Program ICA1-CT-2000-70018 (Center of Excellence CELDIS) and by the U.S. DARPA SpinS Program. One of us (T.W.) was supported by the Fulbright Foundation as a Senior Fulbright Fellow.

## FIGURE CAPTIONS

**Fig. 1.** Hall voltage as a function of magnetic field for two samples of $Ga_{1-x}Mn_xAs$ with $x$ = 0.083: as-grown and annealed at optimal conditions, measured at T = 4.2K. Note that the dominant contribution to the Hall voltage comes from the AHE term. The Hall voltage thus primarily reflects M *vs.* B behavior.

**Fig. 2.** Magnetoresistivity ($R-R_0/R_0$, where $R_0$ is the value of resistivity at B=0) for two epilayers with high Mn concentrations ($x$ = 0.084 and $x$ = 0.081), as-grown and annealed at an optimal temperature (around $280^0$C). Note the distinct decrease of magnetoresistivity observed after heat treatment.

**Fig. 3.** Magnetoresistivity R for a sample with $x$ = 0.01 in low magnetic fields at T = 20K. The magnetic field B was applied perpendicular to the film. Note that a hysteretic behavior is clearly visible. The arrows and numbers indicate the history and direction of applied magnetic field.

**Fig. 4.** Magnetization M(B) for two samples: a) x=0.061, and b) x=0.081 before and after annealing at the optimal temperature of $280^0$C. The magnetic field was applied in the plane of the film. Note that the coercive field of the hysteresis loop decreases after annealing.

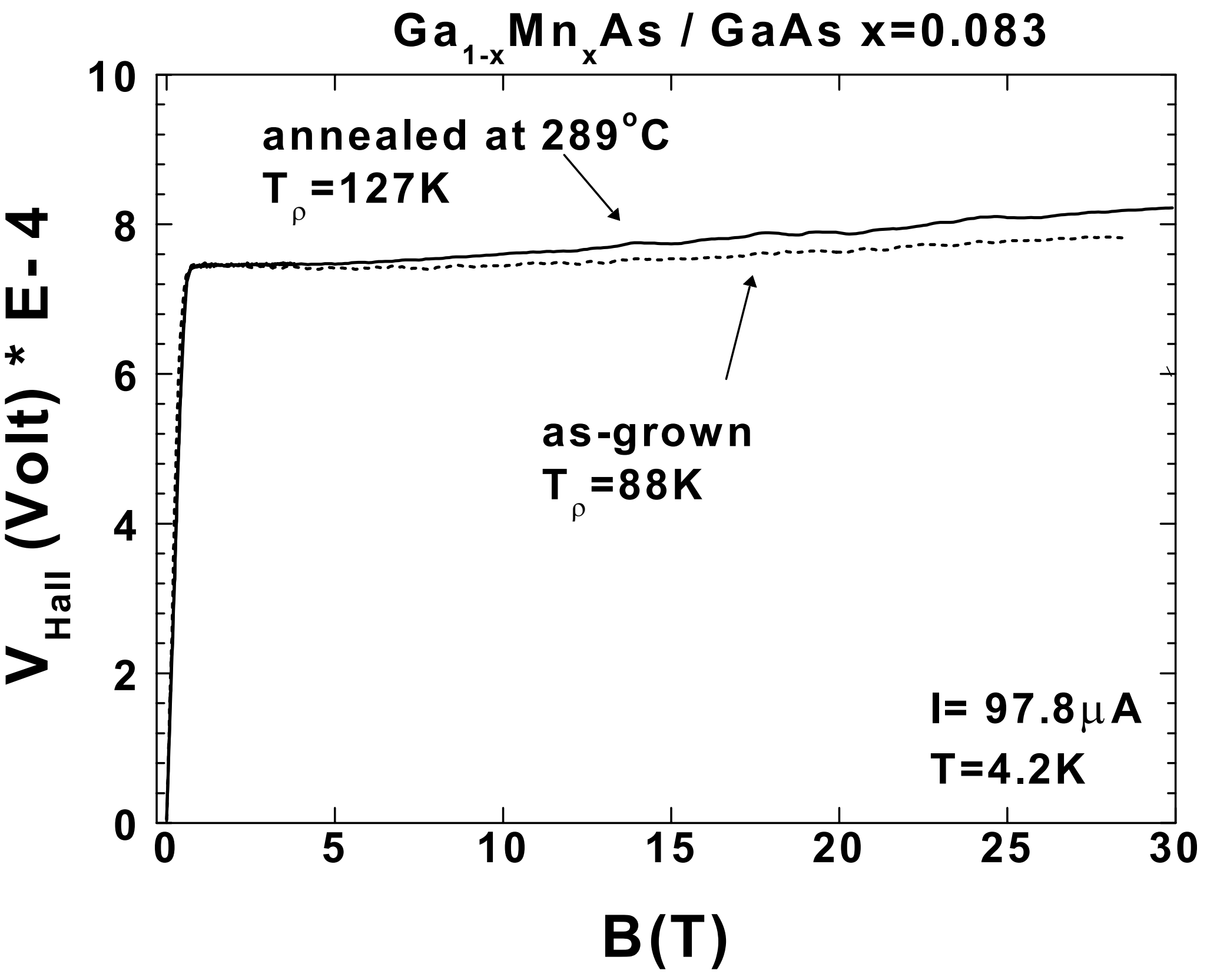

Fig. 1

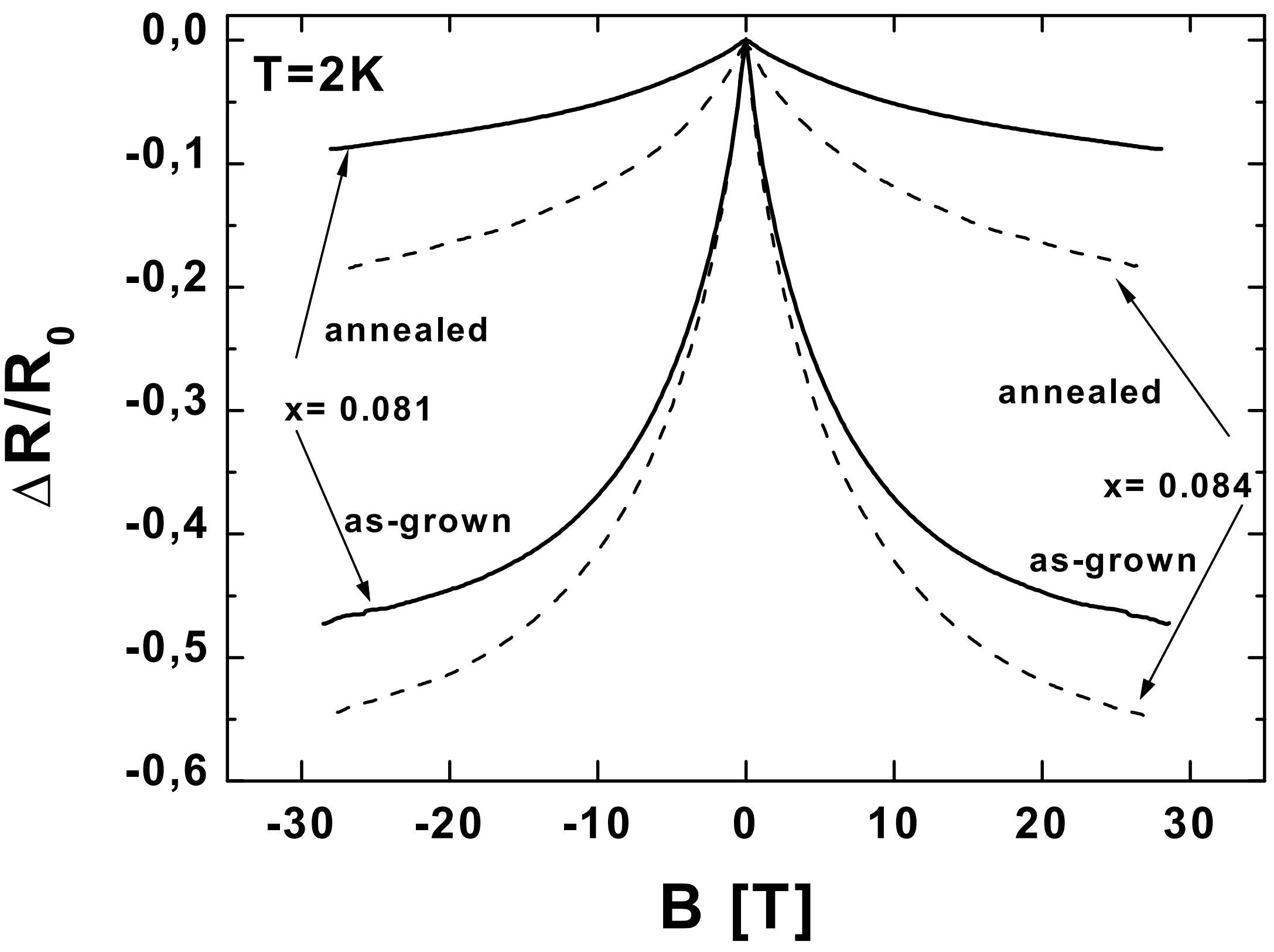

**Fig. 2**

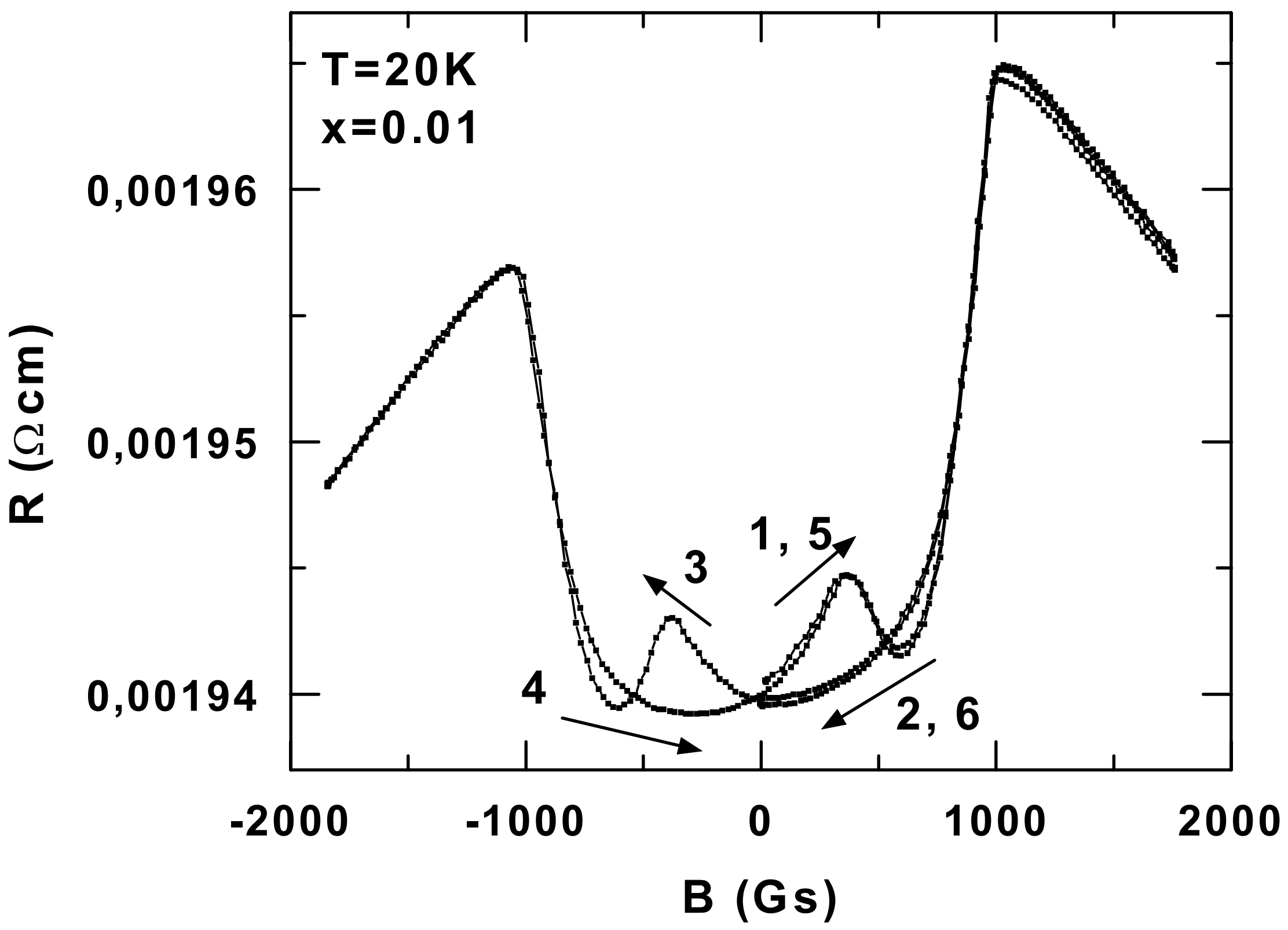

Fig. 3

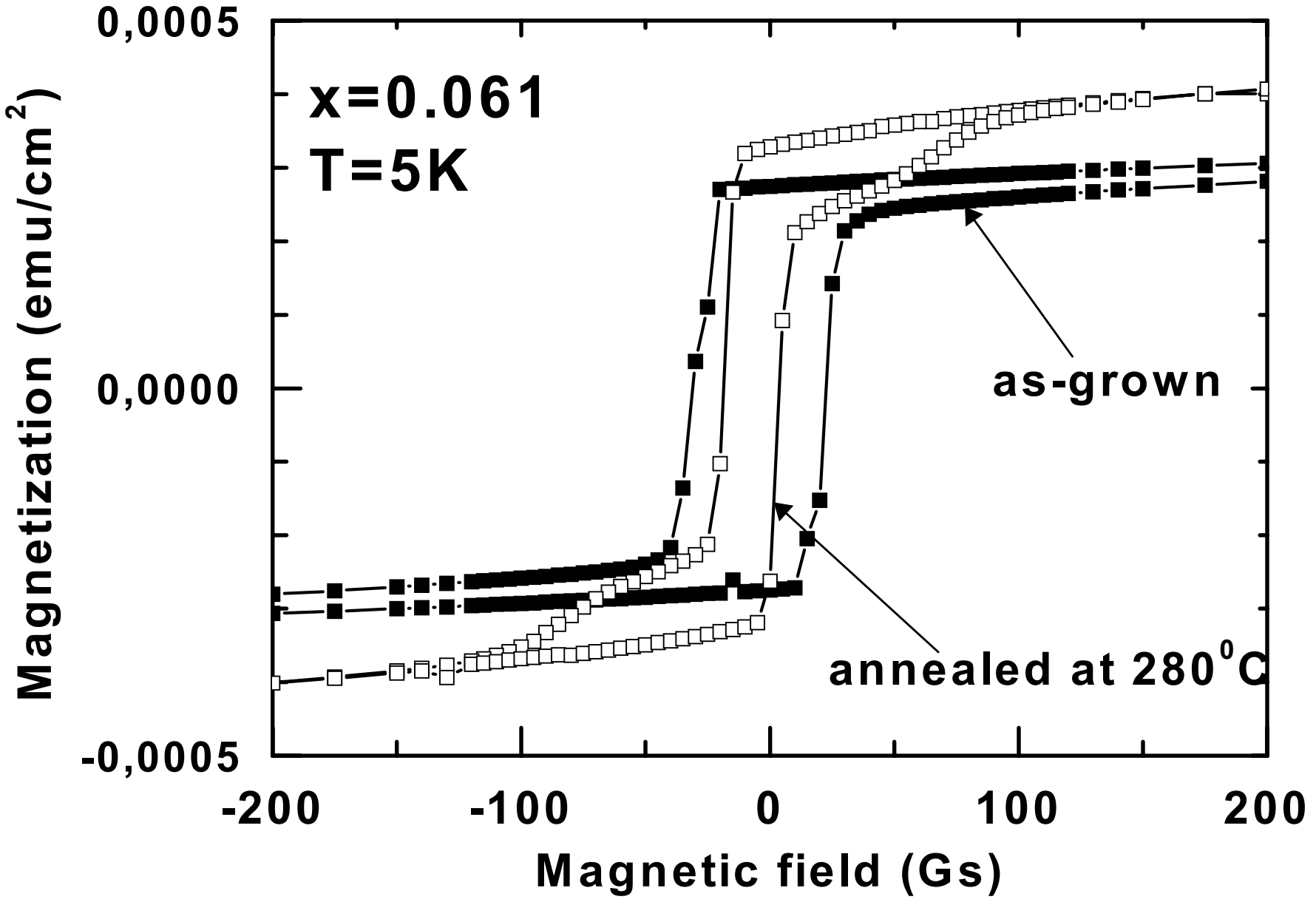

**Fig. 4a)**

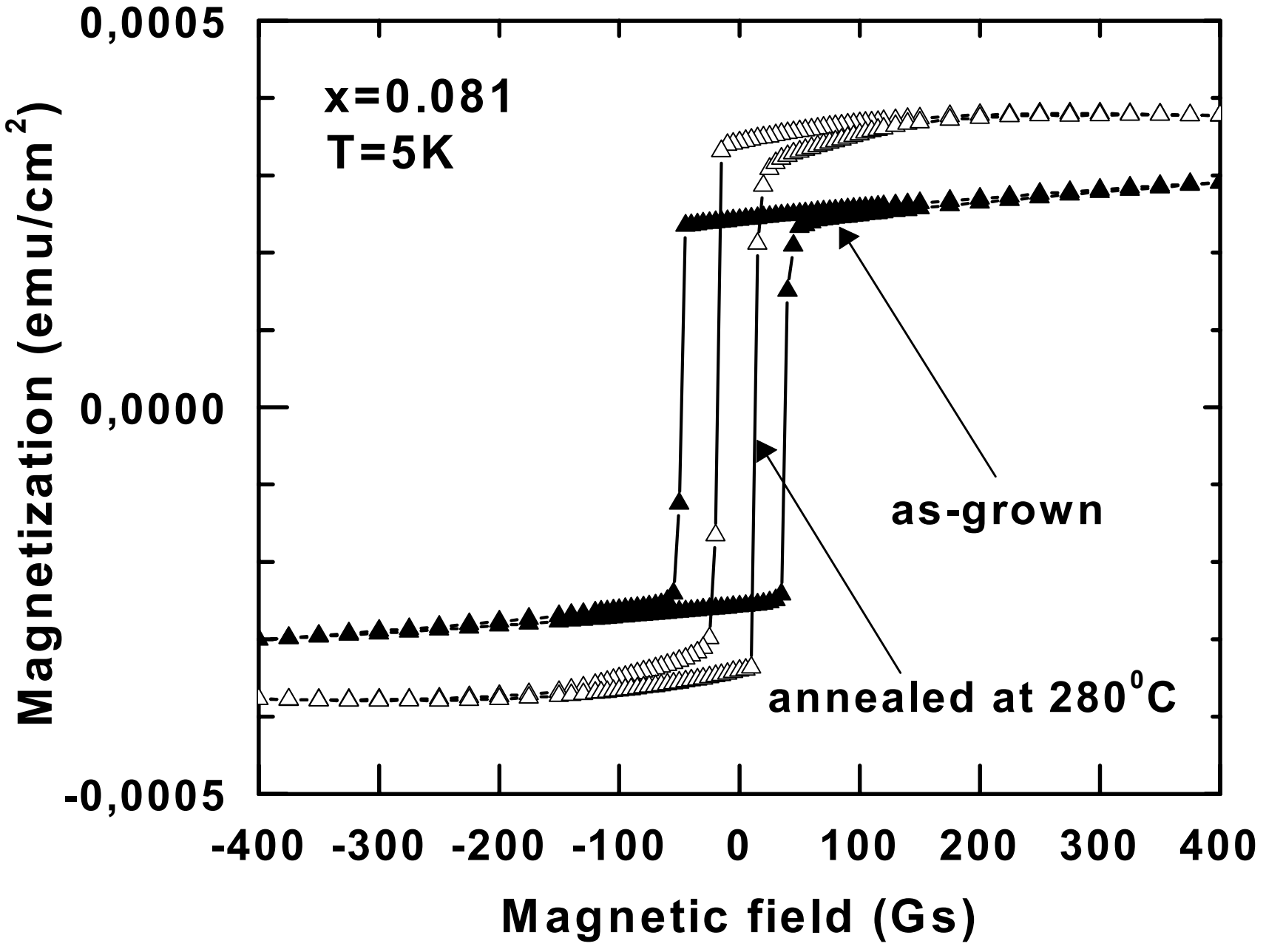

**Fig. 4b)**